\documentstyle[amssymb,epsf,cite,11pt]{article}

\def\baselinestretch{1.3}

\newcommand{\be}{\begin{equation}}
\newcommand{\ee}{\end{equation}}

\begin{document}

\title{\bf \Large Multifractal analysis of weighted networks by a modified sandbox algorithm}

\author{Yu-Qin Song$^{1,2}$, Jin-Long Liu$^{1}$\thanks{Joint first author.}, Zu-Guo Yu$^{1,3}$\thanks{
  Corresponding author, email: yuzuguo@aliyun.com} and Bao-Gen Li$^{1}$  \\
  {\small $^{1}$Hunan Key Laboratory for Computation and Simulation in Science and Engineering and Key Laboratory} \\
  {\small  of Intelligent Computing and Information Processing of Ministry of Education, Xiangtan University, }\\
   {\small  Hunan 411105, China,} \\
  {\small $^2$College of Science, Hunan University of technology, Hunan 412007, China,}\\
{\small $^{3}$School of Mathematical Sciences, Queensland University of Technology, GPO Box 2434,}\\
{\small  Brisbane, Q4001, Australia.}}

\date{}
\maketitle

\begin{abstract}
Complex networks have attracted growing attention in many
fields. As a generalization of fractal analysis, multifractal analysis (MFA)
is a useful way to systematically describe the spatial heterogeneity of
both theoretical and experimental fractal patterns. Some algorithms for MFA of unweighted complex networks
have been proposed in the past a few years, including the sandbox (SB) algorithm recently employed by our group.
In this paper, a modified SB algorithm (we call it SBw algorithm) is proposed for MFA
of weighted networks. First, we use the SBw algorithm to study the multifractal property of two families of
weighted fractal networks (WFNs): ``Sierpinski" WFNs and
 ``Cantor dust" WFNs. We also discuss how the fractal dimension
and generalized fractal dimensions change with the edge-weights of the WFN. From the comparison between
the theoretical and numerical fractal dimensions of these networks, we can find that
the proposed SBw algorithm is efficient and feasible for MFA of weighted
networks. Then, we apply the SBw algorithm to study
multifractal properties of some real weighted networks ---
collaboration networks. It is found that the multifractality exists in these weighted networks, and is affected by their edge-weights.
\end{abstract}

{\bf Key words}: Modified sandbox algorithm; multifractal  analysis; weighted network,
fractal dimension; collaboration network.

\section{Introduction}

~~~ Complex networks have attracted growing attention in many
fields. More and more research works have shown
that they connect with many real complex systems and can be used
in various fields \cite{Song2005,Newman2009,Vidal2011,Xu2015}.
Fundamental properties of complex networks, such as the
small-world, the scale free and communities, have been studied
\cite{Watts2004,Barabasi1999}.
Song {\it et al.} \cite{Song2005} found the self-similarity property \cite{Mandelbrot1983,Feder1988,Falconer1997} of
complex networks. Gallos {\it et al.} gave a review of fractality and self-similarity in complex networks \cite{Gallos2007-1}. At the same time, some methods  for fractal
analysis and how to numerically calculate the fractal dimension of complex networks have been
proposed. Especially, the box-counting algorithm \cite{Song2006,Song2007} was generalized
and applied to calculate the fractal dimension of complex networks.
Subsequently, an improved  algorithm was
proposed to investigate the fractal scaling property in scale-free networks
\cite{Rim2007}. In addition, based on the edge-covering box
counting, an algorithm was proposed to explore the self-similarity
of complex cellular network \cite{Zhou2007}. A ball-covering
approach and an approach defined by the scaling property of the
volume were proposed to calculate the fractal dimension of complex
networks \cite{Gao2008}. Later on, box-covering algorithms for
complex networks were further studied
\cite{NG2011,Schneider2012}.

 Although fractal analysis can
describe global properties of complex networks, it is inadequate
to describe the complexity of complex networks by a single fractal
dimension. For systematically characterizing the spatial
heterogeneity of a fractal object, multifractal analysis (MFA) has
been introduced \cite{Grassberger1983,Halsey1986}. MFA has been widely applied in many fields, such as
financial modeling \cite{Canessa2000,Anh2000}, biological systems
\cite{Yu2001a,Yu2001b,Anh2002,Yu2003,Yu2004,Zhou2005,Yu2006,Yu2010b,Han2010,Zhu2011,Zhou2014},
geophysical systems
\cite{Kantelhardt2006,Veneziano2006,Venugopal2006,Yu2007,Zang2007,Yu2009,Yu2010a,Yu2014}
and also complex networks \cite{Furuya2011,Wang2012,Li2014,Liu2014,Liu2015}.
Lee {\it et al.} \cite{Lee2006} mentioned that MFA is the best tool to describe the
probability distribution of the clustering coefficient of a
complex network. Some
algorithms were proposed for MFA of unweighted complex networks in past a few years
\cite{Furuya2011,Wang2012,Li2014,Liu2014,Liu2015}. Furuya and  Yakubo
\cite{Furuya2011} pointed out that a single fractal dimension is not enough to
characterize  the fractal property of a scale-free network  when
the network has a multifractal structure. They also introduced a compact-box-burning
(CBB) algorithm for MFA of complex networks. Wang {\it et al.}
\cite{Wang2012} proposed an improved fixed-size box-counting
algorithm to study the multifractal behavior of complex networks.
Then this algorithm was improved further by Li {\it et al.} \cite{Li2014}.
They applied the improved fixed-size box-counting algorithm to study
multifractal properties of a family of fractal
networks proposed by Gallos {\it et al.} \cite{Gallos2007}.
Recently,  Liu {\it et al.} \cite{Liu2015} employed the sandbox (SB)
algorithm proposed by T\'{e}l {\it et al.} \cite{Tel1989} for
MFA of complex networks. The comparison between theoretical and numerical
results of some networks showed
that the SB algorithm is the most effective and feasible
algorithm to study the multifractal behavior of unweighted  networks \cite{Liu2015}.

  However, all the algorithms for MFA in Refs. [41-45] are just feasible
for unweighted networks. Actually, there are many weighted networks
in real world \cite{Bagler2008,Hwang2010,Cai2012}, but few works
have been done to study the fractal and multifractal properties of the weighted
networks.  Recently, an improved box-covering
algorithm for weighted networks was proposed by Wei {\it et al.}
\cite{Wei2013}. They applied the box-covering algorithm for weighted
networks (BCANw) to calculate the fractal dimension of the
``Sierpinski" weighted fractal network (WFN) \cite{Carletti2010}
and some real weighted networks. But the BCANw algorithm was only designed
 for calculating the fractal dimension of weighted networks.

 In this work, motivated by the idea of BCANw, we propose a modified sandbox algorithm (we call it SBw algorithm) for MFA of weighted networks. First, we use the SBw algorithm to study the multifractal property of two families of
weighted fractal networks (WFNs): ``Sierpinski" WFNs and
 ``Cantor dust" WFNs introduced by Carletti {\it et al.} \cite{Carletti2010}. We also discuss how the fractal dimension and generalized fractal dimensions change with the edge-weights of the WFN. Through the comparison between
the theoretical and numerical fractal dimensions of these networks, we check whether
the proposed SBw algorithm is efficient and feasible for MFA of weighted
networks. Then, we apply the SBw algorithm to study
multifractal properties of some real weighted networks ---
collaboration networks \cite{Newman2001}.

\section*{Results and Discussion}

\subsection*{Multifractal properties of two families of weighted fractal networks }
\begin{figure}[ht]
\centerline{\epsfxsize=11cm \epsfbox{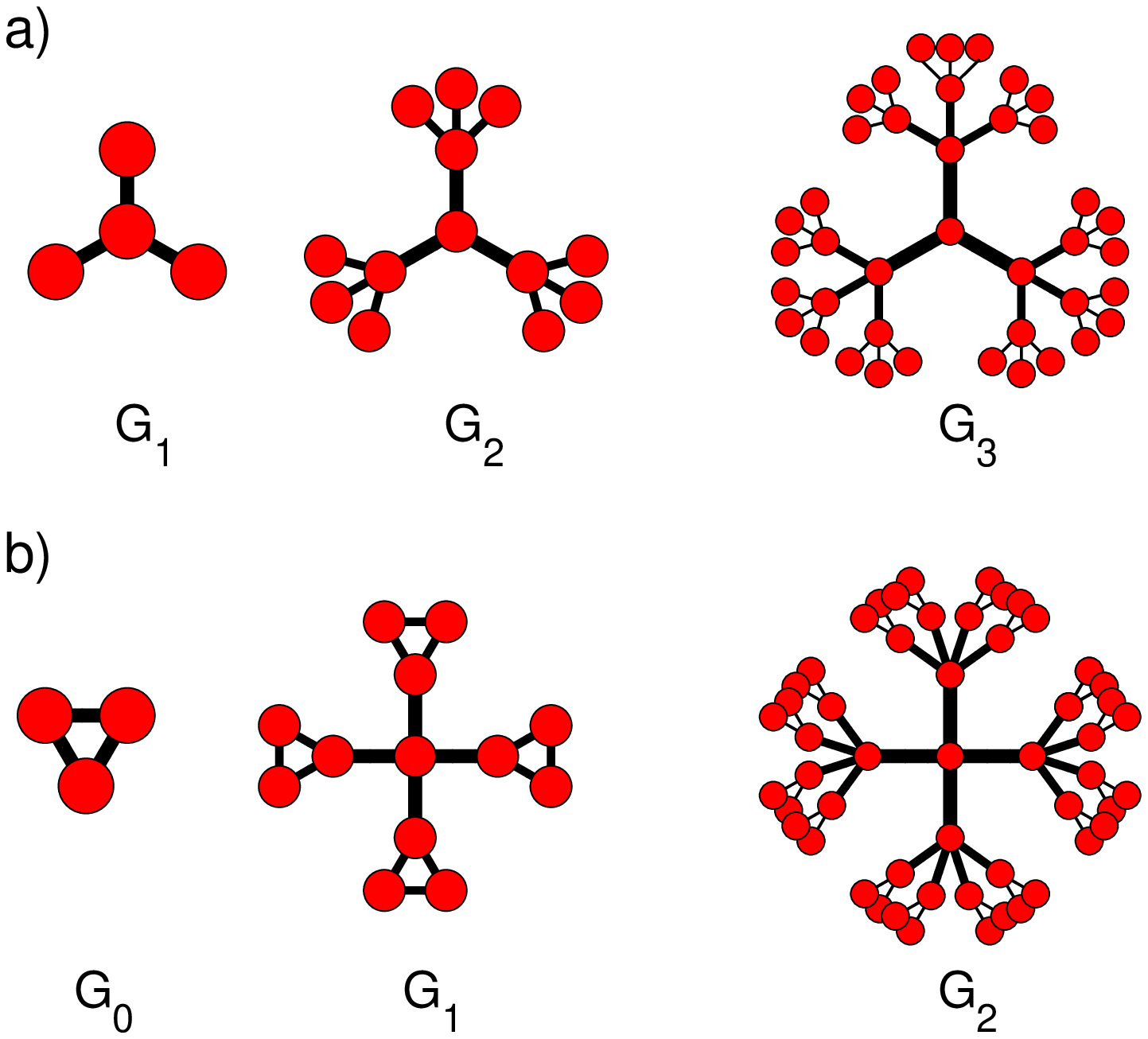}}
\caption{{\bf a)} The ``Sierpinski" weighted fractal networks, s=3, f=1/2
  and $ G_{0} $ is composed by a single node. From the left to the
  right, the 1th generation $ G_{1} $, the 2th generation $ G_{2}$, and the 3th
generation $ G_{3} $ are shown.  The fractal dimension of the limit network
is $\log(3)/\log(2)\approx 1.5850$. {\bf b)} The ``Cantor dust" weighted fractal networks, s=4, f=1/5
  and $G_{0}$ is a triangle. From the left to the
  right, the 0th generation $ G_{0}$, the 1th generation $G_{1}$, and the 2th
generation $G_{2}$ are shown.  The fractal dimension of the limit network
is $\log(4)/\log(5)\approx 0.8614$.}
\label{fig:1}
\end{figure}

~~~  In order to show that the SBw algorithm for MFA of
 weighted network is effective and feasible, we
 apply our method to study the multifractal behavior of  the ``Sierpinski"
WFNs and the ``Cantor dust"  WFNs \cite{Carletti2010}. These WFNs are
constructed by Iterated Function Systems
 (IFS) \cite{Barnsley2001}, whose Hausdorff dimension
 is completely characterized by two main parameters: the number
 of copies $s>1$ and the scaling factor $0<f<1$ of the IFS. In this
 case, the fractal dimension of the fractal weighted network is
 called the similarity dimension and given by \cite{Carletti2010}:
\begin{equation}
   d_{fract}=-\frac{\log(s)}{\log(f)}.
\end{equation}
To construct ``Sierpinski" WFNs and ``Cantor
dust" WFNs \cite{Carletti2010}, a single node and a triangle is set as a
initial network $G_{0}$ respectively. The first a few steps to construct them
are shown in parts a) and b) of Figure \ref{fig:1} respectively.

  We first consider two ``Sierpinski" WFNs  with
parameters $s=3, f=1/2 $ and $ s=3,f=1/3 $ respectively. Considering the limitation of the
computing capability of our computer, we construct the 8th
generation $G_{8}$ of these two networks. There are 9841 nodes and 9837
edges in the $G_{8}$ of these two networks. For the case $s=3, f=1/2 $, the edge-weights
of $G_{8}$ are equal to $1,
1/2, 1/4,1/8,1/16,1/32, 1/64,1/128$, respectively; the diameter of
$G_{8}$ is less than 4. When we use the SBw algorithm for MFA of
$G_{8}$, radiuses  $r$  of sandboxes are
set to $1/128, 1/128+1/64, \cdots, 1+1/2+
1/4+1/8+1/16+1/32+1/64+1/128 $, respectively for this case. We can do similar analysis for
$G_{8}$ of network with $ s=3,f=1/3 $.
 It is an important step to look for an appropriate range of
$r~(i.e.,r\in[r_{min},r_{max}]) $ for obtaining the generalized
fractal dimensions $D(q)$ (defined by equations (6) and (7)) and the mass exponents $\tau(q)$ (defined by equation (5)). In this
paper, we set the range of $ q $ values from  $-10 $  to $ 10
$ with a step of $1$.

\begin{figure}[ht]
\centerline{\epsfxsize=11cm \epsfbox{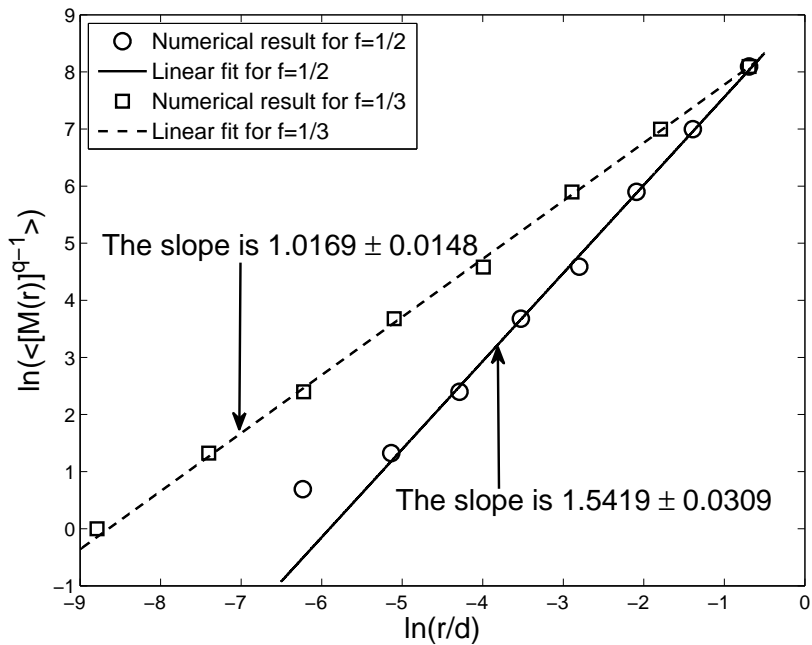}}
\caption{Examples of fractal analysis of the
   "Sierpinski" weighted fractal networks $G_{8}$ with $9841$
nodes. Here, copy factor $s=3$ and the scaling factor $f=1/2,
1/3$, respectively. By means of the least square fit, the slope
of the reference lines are $1.5419\pm 0.0309$ and $1.0169\pm 0.0148$ respectively.
The theoretical result is $1.5850$ (for $f=1/2$) and 1.0 (for $f=1/3$),
respectively.}
\label{fig:2}
\end{figure}

   When $ q=0 $,  $ D(0) $ is
the fractal dimension of a complex network.
Now we adopt the SBw algorithm to estimate the fractal dimension of two ``Sierpinski" WFNs  with
parameters $s=3, f=1/2 $ and $ s=3,f=1/3 $ respectively.
We show the linear regression of $ \ln(\langle [M(r)]^{q-1}\rangle)$ against
$(q-1)\ln(r/d)$ for $ q=0 $ in Figure \ref{fig:2}. By
means of the least square fit, the slope of the reference lines are estimated
to be $ 1.5419 $ and $ 1.0169 $, with standard deviations 0.0309 and 0.0148, respectively.
It means that the numerical fractal dimension is $1.5419\pm 0.0309 $ and $ 1.0169\pm 0.0148 $,
respectively; they are very close to the theoretical similarity dimension
$1.5850 $ and $ 1.0 $ respectively. Hence we can say that the numerical
 fractal dimension obtained by the SBw algorithm is very close to the theoretical
similarity dimension for a ``Sierpinski" WFN.

   To further check the validity of the SBw algorithm,
   let the copy factor $ s $ be $ 3 $ and the scaling factor $ f $ be
any  positive real number in the range  $ 0<f<1 $. From Equation
(1), we can get the relationship between the fractal dimension and
the scaling factor $ f $ of the ``Sierpinski" WFN  as:
\begin{equation}
   d_{fract}=-\frac{\log(3)}{\log(f)}.
\end{equation}

\begin{figure}[ht]
\centerline{\epsfxsize=11cm \epsfbox{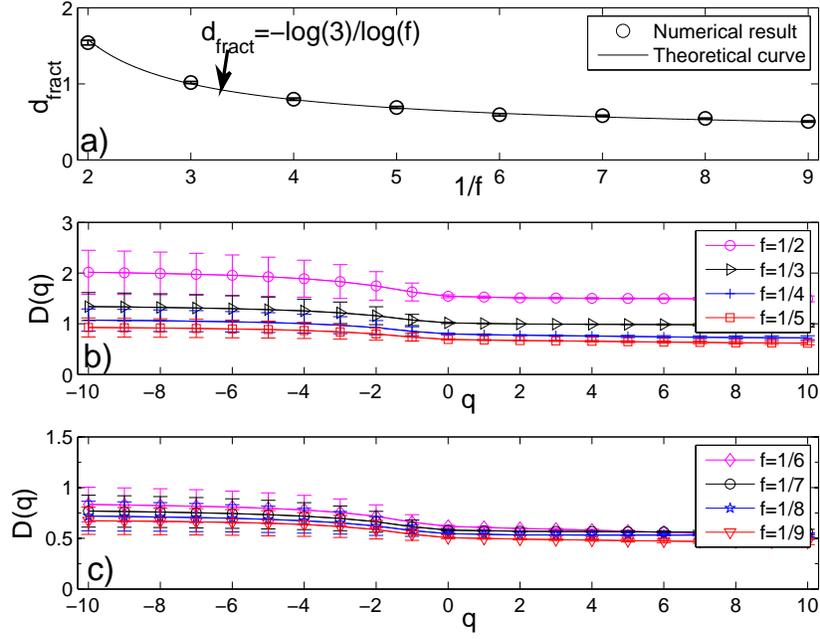}}
\caption{{\bf a)} The fractal dimensions  and their standard deviations of  $G_{8}$ of ``Sierpinski" WFNs with parameter $ s=3$. The solid curve represents the theoretical $d _{fract}$ given by Eq. (2), circles are the numerical fractal dimensions estimated
  by the SBw algorithm.  {\bf b) and c) } The generalized fractal dimensions $D(q)$ curves and their standard deviations of the 8th generation $G_{8}$ of ``Sierpinski" WFNs
    estimated by the SBw algorithm. Here, the parameter $ s=3$,
   $ f=1/2,1/3,1/4,1/5$ and $1/6,1/7,1/8,1/9 $, respectively. Each error bar takes twice length to the standard deviation for all the results.}
\label{fig:3}
\end{figure}

  For each value of $ f=1/2, 1/3, 1/4, 1/5, 1/6, 1/7, 1/8, 1/9 $,
  we calculate fractal dimensions and their standard deviations
 of the 8th generation ``Sierpinski"  WFN $ G_{8} $ by the SBw algorithm.
 The results are shown in part a) of Figure \ref{fig:3}, where each error bar takes twice length to the standard deviation.
 This figure shows that the numerical fractal dimensions obtained by the SBw algorithm agree well with the
  theoretical fractal dimensions of these networks. This figure also shows that
  the fractal dimension of WFNs is affected by the edge-weight.
  This result coincides with the conclusion obtained by Wei {\it et al.}\cite{Wei2013}.

   Hence we can apply the SBw algorithm to calculate the generalized
fractal dimensions $D(q)$ and their standard deviations of ``Sierpinski" WFNs.
In parts b) and c) of Figure \ref{fig:3}, we show the  generalized
fractal dimensions $D(q)$ of the 8th generation $G_{8}$ of ``Sierpinski" WFNs,
 with the parameter $s=3$, $f=1/2,1/3,1/4,1/5$ and $1/6,1/7,1/8,1/9 $ respectively.
From these figures, we can see that all the 8th generation $G_{8}$ of ``Sierpinski" WFNs for
different $f$ have multifractal property, and the
multifractal property of these weighted networks is affected by
their edge-weights. The result also shows that the generalized fractal
dimension $D(q)$ almost decreases with the decrease of the scaling factor $ f $ for any $q$.

\begin{figure}[ht]
\centerline{\epsfxsize=11cm \epsfbox{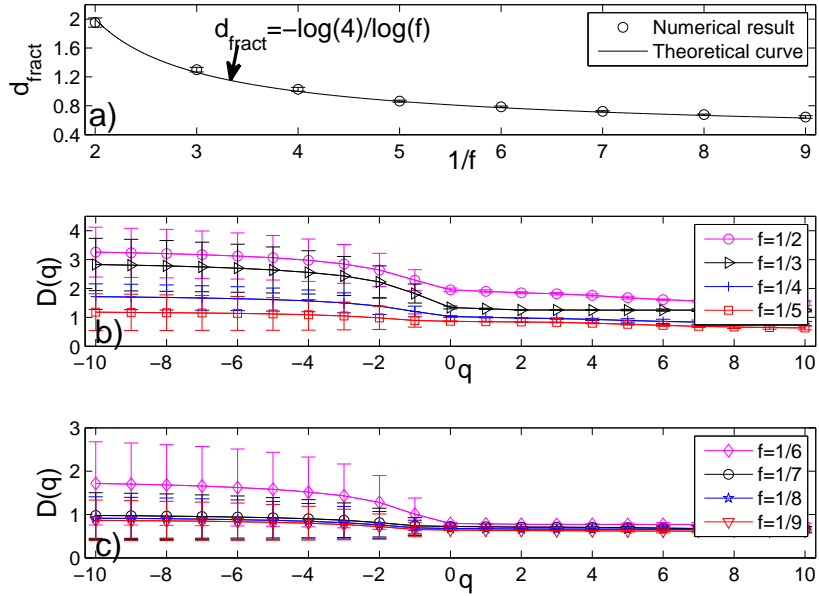}}
\caption{{\bf a)} The fractal dimensions and their standard deviations of $G_{5}$ of ``Cantor dust" WFNs
    with parameter $s=4$. The solid curve
  represent the theoretical $d _{fract}$ given by Eq. (1),
  circles indicate the numerical fractal dimension estimated
  by the SBw algorithm. {\bf b) and c) } The generalized fractal dimensions $D(q)$ curves and their standard deviations of $G_{5}$ of ``Cantor dust" WFNs
  estimated by the SBw algorithm. Here, the parameter $ s=4 $,
   $ f=1/2,1/3,1/4,1/5$ and $1/6,1/7,1/8,1/9 $, respectively. Each error bar takes twice length to the standard deviation for all the results.}
\label{fig:4}
\end{figure}

 For ``Cantor dust"  WFNs,  we can only construct the 5th
  generation networks with $s=4$ and
  $f=1/2,1/3,1/4$, $1/5,1/6,1/7$, $1/8,1/9$, respectively.
We first calculate fractal dimensions and their standard deviations of these WFNs by the
SBw algorithm. The results are shown in part a) of Figure \ref{fig:4}.
From this figure, we can see that the
numerical fractal dimensions obtained by the SBw algorithm are very
close to the theoretical fractal dimensions $d_{fract}=-\log(4)/\log(f)$ for these WFNs.
   Then we apply the SBw algorithm to calculate the generalized
fractal dimensions $D(q)$ and their standard deviations of these ``Cantor dust"  WFNs. We show the numerical results of
the 5th generation  $G_{5}$ of ``Cantor dust"  WFNs in parts b) and c) of Figure \ref{fig:4}. From
these figures, we can see that all $D(q)$ curves are nonlinear. It indicates
that all these weighted networks have multifractal property.
Similar to "Sierpinski" WFNs, the multifractal property of
these networks is affected by their edge-weights.

The multifractal property of ``Sierpinski"
WFNs and ``Cantor dust"  WFNs
revealed by the SBw algorithm indicates that these model networks are
very complicated, and cannot be characterized by a single fractal
dimension.

\subsection*{Applications: multifractal properties of three collaboration networks}

~~~  Now we apply the SBw algorithm to study multifractal properties of some real networks. We study
three collaboration networks: the high-energy theory collaboration network
\cite{Newman2001}, the astrophysics collaboration network \cite{Newman2001},
and the computational geometry collaboration
network \cite{Data}.

\begin{figure}[ht]
\centerline{\epsfxsize=11cm \epsfbox{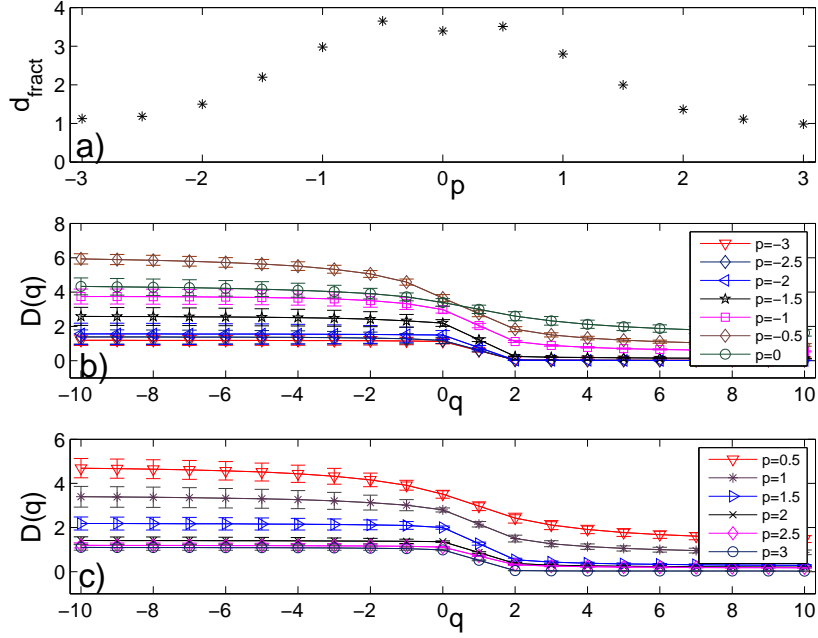}}
\caption{{\bf a)} The relation between values of the fractal dimension
  of the High-energy theory collaboration networks and values of $p$.
  We set the range of the $ p $ values from  $-3 $  to $ 3 $ with a step of
 $0.5$. {\bf b) and c)} The generalized fractal dimensions $D(q)$ curves and their standard deviations  of the
  the High-energy theory collaboration network  by using the
   SBw algorithm. Here, the range of the $ p $ values from  $-3 $
    to $ 3 $ with a step of $0.5$. Each error bar takes twice length to the standard deviation for all the results.}
\label{fig:5}
\end{figure}

{\bf High-energy theory collaboration network:}
 This network has $8361$ nodes and $15751$ edges, the edge-weights are defined as \cite{Newman2001}:
\begin{equation}
   w_{ij}=\sum_{k}\frac{\delta ^{k} _{i}\delta ^{k}
   _{j}}{n_{k}-1},
\end{equation}
where $n_{k}$ is the number of co-author in the  $k$th paper
(excluding single authored papers), $\delta ^{k} _{i}$  equals to $1$
if the $i$th scientist is one of the co-author of the $k$th paper,
otherwise it equals to $0$. The data contains all components
of the network, for a total of 8361 scientists, not just the
largest component of 5835 scientists. When two authors share many
papers, the weight value is larger, thus the distance is less. So,
in Equation(9), $p$ had better be a negative number (e.g. $-1$ given by Newman
\cite{Newman2001}). For different values of
$p$, we can calculate the shortest path by Equation(9) and obtain
different weighted networks. Then we apply the SBw algorithm to
calculate the generalized fractal dimensions $D(q)$ and their standard deviations of the largest
component of the network with 5835 nodes. We show the
relation between the numerical fractal dimension of the
High-energy theory collaboration networks and values of $p$ in part a) of Figure \ref{fig:5}.
From this figure, we can see the value of fractal dimension decreases with the
increase of the absolute value of $p$, the values of fractal
dimensions are almost symmetric about the vertical axis.
 We show the numerical results on the
generalized fractal dimensions $D(q)$ of the High-energy theory
collaboration networks for different values of $p$ in parts b) and c) of Figure \ref{fig:5}.
From these figures, we can see that all the High-energy theory collaboration
networks for different  $p$ have multifractal property, and the multifractal property of these weighted networks
is affected by the edge-weight. We can also see that the
generalized fractal dimensions $D(q)$ almost decrease  with the
increase of the absolute value of $p$.

\begin{figure}[ht]
\centerline{\epsfxsize=11cm \epsfbox{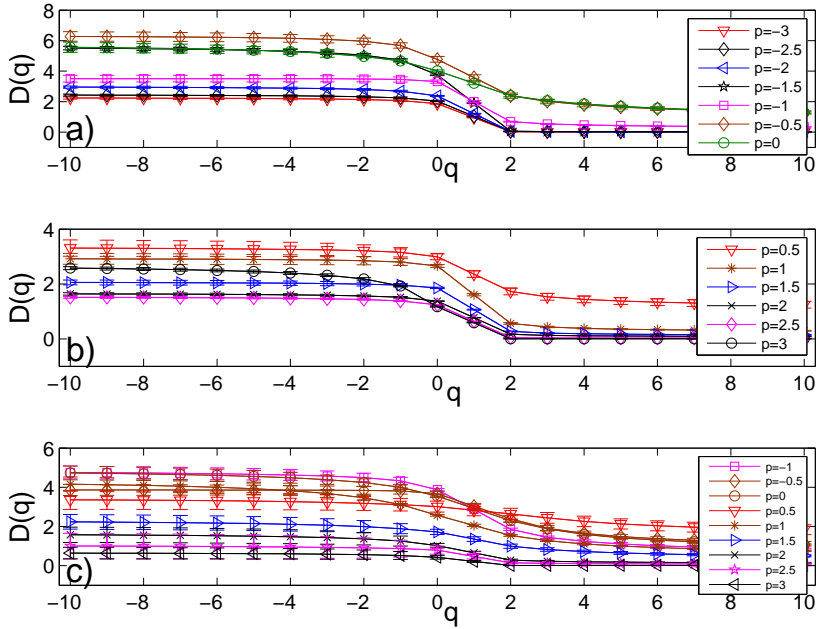}}
\caption{ The generalized fractal dimensions $D(q)$ curves and their standard deviations of
  {\bf a) and b)} the astrophysics collaboration networks, and {\bf c)} the computational geometry collaboration networks estimated  by the
   SBw algorithm. Here, we set the range of the $ p $ values from  $-1 $  to $ 3 $ with a step of
 $0.5$. Each error bar takes twice length to the standard deviation for all the results.}
\label{fig:6}
\end{figure}

{\bf Astrophysics collaboration network: }
This network has $16706$ nodes and $121251$ edges, the edge-wights is defined as
 Equation(3). Here, the data contains all components of the network,
 for a total of $16706$ scientists, not just the largest component of
$14845$ scientists. When two authors share many papers, the weight
value is larger, thus the distance is less. So, in  Equation(9),
$p$ had also better be a negative number (e.g. $-1$ given by Newman
\cite{Newman2001}). We calculate the shortest
path by Equation (9) and obtain some weighted networks with different values of $p$. Then we apply the SBw
algorithm to calculate the generalized fractal dimensions $D(q)$ and their standard deviations of the largest component of the network with $14845$ nodes.
We show the numerical results of the astrophysics collaboration networks in parts a)  and b) of Figure \ref{fig:6}.
From this figure, we can see that these networks also have
multifractal property, and the multifractal property
of these weighted networks is affected by the edge-weight.

{\bf Computational geometry collaboration network: }
 The authors collaboration network in computational
geometry was produced from the BibTeX bibliography which obtained
from the Computational Geometry Database. This network has 7343
nodes and 11898 edges. Two authors are linked with an edge, if and
only if they wrote a common paper or book, etc. The value of
edge-weight is the number of common works, so the value is one integer,
 such as $1,2,3,\cdots$, etc. The data contains all
components of the network, for a total of 7343 scientists, not
just the largest component of 3621 scientists. The data can be got
from Pajek Data \cite{Data}. When two authors share many papers,
the weight value is larger, thus the distance is less. So, in Equation
(9), $p$ had better be a negative number. We calculate the shortest
path by Equation (9) and obtain some weighted networks with different values of $p$. Then we apply the SBw
algorithm to calculate the generalized fractal dimensions $D(q)$ and their standard deviations
of the largest component of the network with 3621 nodes. Because the way to define the weight of this network is different from another two real networks, we can only calculate the generalized fractal dimensions $D(q)$ and their standard deviations
of the largest component of the network with 3621 nodes for $p\ge -1$. We show
the numerical results of the computational geometry collaboration
networks in part c) of Figure \ref{fig:6}. From this figure, we can also see that these networks have multifractal property, and the
multifractal property of these weighted networks is affected by
the edge-weight (but the impact is relatively small).

\section*{Conclusions}

~~~ In this paper, a modified sandbox algorithm (we call it SBw algorithm) for MFA
of weighted networks is proposed. First, we used the SBw algorithm to study the multifractal property of two families of weighted fractal networks (WFNs): ``Sierpinski" WFNs and ``Cantor dust" WFNs. We also discussed how the fractal dimension and generalized fractal dimensions change with the edge-weights of the WFN. From the comparison between
the theoretical and numerical fractal dimensions of these networks, we can find that
the proposed SBw algorithm is efficient and feasible for MFA of weighted
networks.

In addition, we applied the SBw algorithm
to study the multifractal properties of some real networks --- the high-energy theory collaboration network, the astrophysics
collaboration network, and the computational geometry
collaboration network. We found that multifractality exists
in these weighted networks, and  is also affected by their
edge-weight. Our result indicates that multifractal property of
weighted networks are affected both by their edge weight and
their topology structure.

\section*{Methods}

\subsection*{ Multifractal analysis}

 ~~~ The fixed-size box-counting
 algorithm is one of the most common and effective algorithms to explore
 multifractal properties of fractal sets\cite{Halsey1986}.
 For a support set $E$ in a metric space $\Omega$ and a normalized measure $\mu$ (i.e. $0\leq\mu(\Omega)\leq 1$),
 we consider the partition sum:
\begin{equation}
Z_{\varepsilon}(q)=\sum\limits_{\mu(B)}[\mu(B)]^{q},
\end{equation}
where $q\in R$, and the sum runs over all different
non-overlapping boxes $ B $ which cover the support set $E$ with a
given size $\varepsilon$.  The mass
exponents $\tau(q)$ of the measure $\mu$ is defined as:
 \begin{equation}
    \tau(q)=\lim\limits_{\varepsilon\rightarrow 0} \frac{\ln Z_{\varepsilon}(q)}{\ln
    \varepsilon}.
\end{equation}
   The generalized fractal dimension $D(q)$ of the measure  $\mu$ is defined as:
\begin{equation}
 D(q)=\left \{\begin{array} {c@{\quad}c}
 \frac{\tau(q)}{q-1}, & q\neq 1,\\
\lim\limits_{\varepsilon\rightarrow
0}{\frac{Z_{1,\varepsilon}}{\ln \varepsilon}}, & q=1.
\end{array}\right.
\end{equation}
where $ Z_{1,\varepsilon}= \sum\limits_{\mu(b)\neq
0}\mu(B)\ln\mu(B)$.  A numerical estimation of the generalized
fractal dimension $D(q)$ can be got from the linear regression of
$ \ln Z_{\varepsilon}(q)/q-1 $ against  $\ln \varepsilon$ for
$q\neq 1$, $Z_{1,\varepsilon}$ against  $\ln \varepsilon$ for $q=
1$, respectively.

T\`{e}l {\it et al.} \cite{Tel1989} proposed the sandbox (SB) algorithm for MFA of fractal sets which
is an extension of the box-counting algorithm \cite{Halsey1986}.
The generalized fractal dimensions $D(q)$  are defined as \cite{Tel1989}:
\begin{equation}
    D(q)=\lim\limits_{r\rightarrow 0} \frac{\ln\langle [M(r)/M(0)]^{q-1} \rangle}
     {\ln(r/d)} \frac{1}{q-1},     ~~~~ q \in R,
\end{equation}
 where $ M(r) $ is the number of points in the sandbox with radius $r$, $M(0)$ is the number
 of all points in the fractal object. It is denoted the brackets $ \langle \cdot \rangle$ to
 take statistical average over randomly chosen centers of the sandboxes. From Equation (7)
 we can get the relation:
\begin{equation}
   \ln(\langle [M(r)]^{q-1}\rangle)\varpropto
   D(q)(q-1)\ln(r/d)+(q-1)\ln (M_{0}).
\end{equation}
 From Equation (8), we can obtain an estimation of the generalized fractal dimension
  $D(q)$  by the linear regression of  $ \ln(\langle [M(r)]^{q-1}\rangle)$ against
  $(q-1)\ln(r/d)$. Then, we can also get the mass exponents $\tau(q)$ through $\tau(q)=(q-1)D(q)$.
   Specifically, $ D(0)$ is the fractal dimension, $ D(1)$ is the information dimension,  $ D(2)$
    is the correlation dimension of the fractal object, respectively.

\subsection*{A modified sandbox algorithm for multifractal analysis of weighted networks}

~~~ Recently,  our group employed the SB
algorithm proposed by T\'{e}l {\it et al.} \cite{Tel1989} for
MFA of unweighted complex networks \cite{Liu2015}. In the SB algorithm \cite{Liu2015}, the radiuses $r$ of the
sandbox are set to be integers in the range from $1$ to the
diameter of the unweighted network. However, in weighted networks, the values
of edge-weights could be any real numbers excluding zero and the
shortest path is defined by the path between two nodes
such that the sum of values of its edge-weights to be minimized in some way
\cite{Newman2004}. So, the shortest path between two nodes could
be any real numbers excluding zero. In this paper, for weighted networks,
we denote the length of shortest path between node $i$ and node $j$
by $d_{ij}$, and $d_{ij}$ is defined as \cite{Wei2013}:
\begin{equation}
   d_{ij}=\min(w_{ij_{1}}^{p}+w_{j_{1}j_{2}}^{p}+\cdots+w_{j_{m-1}j_{m}}^{p}+w_{j_{m}j}^{p}),
\end{equation}
where $ w_{kh} $ means the edge-weight of directly connecting node
$k$ and node $h$ in a path, $j_{m}(m=1,2,\cdots) $ are IDs of nodes and $
p$ is a real number. In particular, when $ p $ equal to zero, the length of the
shortest path given by Equation(9) is the same as unweighted networks
\cite{Newman2004}. If the edge-weight is only a number without
obvious physical meaning, we set $p$ equals to $1$, such as the
``Sierpinski" WFN \cite{Carletti2010}. In some real weighted
networks, one case is that the bigger edge-weight of
between any two nodes is, the less distance is, such as the
collaboration networks, where $ p < 0 $ \cite{Newman2001}; the
other case is that the bigger edge-weight of between any two nodes
is, the further distance is, such as the real city network and the
``Sierpinski" WFN , where $ p> 0 $.

The SB algorithm is unfeasible for MFA of weighted networks because we cannot obtain enough numbers of boxes (even only one
sandbox we can obtain when the diameter of the weighted network is less than one).
Wei {\it et al.} \cite{Wei2013} proposed an improved box-covering
algorithm for fractal analysis of weighted network (BCANw). In the present work, motivated by the idea of BCANw, we propose a modified sandbox algorithm (we call it SBw algorithm) for MFA of weighted networks. The
SBw algorithm can deal with the multifractal property (hence can also deal with the fractal property) of weighted networks.

   Before we apply the SBw algorithm for MFA of weighted networks, we
need to calculate the shortest-path distance matrix $ D $ of the
network and set the range of radiuses $ r $ of the sandboxes. The
detail is given as:
\begin{itemize}

   \item  A network is mapped to an adjacent matrix $ W_{N\times N}$,
    where $ N $ is the total number of nodes in the network. For any given real numbers $ p $, the elements of the
    adjacent matrix $ w_{ij}^{p}\neq 0$ is the edge-weight between directly connecting nodes $ i $ and $ j $,
    otherwise $ w_{ij}^{p}=0$. According to the adjacent
    matrix $ W_{N\times N}$, we can calculate the shortest path
    distance matrix $ D $ by applying the Floyd's
    algorithm \cite{Floyd1962} of Matlab BGL toolbox \cite{Gleich};

    \item  For any given real numbers $ p $, order the edge-weights $ w_{ij}^{p} $ as
    $w_{1}\leq w_{2}\leq\cdots \leq w_{m}$, where $m $ is the number of
    edge-weights. From the fractal theory, we should look for an appropriate range of radiuses $r$ to perform the least square linear fit and then obtain the generalized fractal dimensions $D(q)$ accurately. We tried choosing the radius $r$ from 0 to diameter $d$ with equal (linearly or logarithmically) intervals. But we found it is hard to look for an appropriate range of radiuses $r$ to perform the least square linear fit and then obtain the generalized fractal dimensions $D(q)$ of weighted complex networks we considered accurately. So the radiuses $ r $ of the sandboxes are obtained
    by accumulating the value of the edge-weights until it is larger than the
    diameter $ d $ of the network. So, we can get the set of radiuses (
    denoted as $ R $), where $ R=\{w_{1} , w_{1}+w_{2},\cdots, \sum_{i=1}^{k}w_{i}:\ k\leq m\}$
    and  $\sum_{i=1}^{K}w_{i}\leq d < \sum_{i=1}^{K+1}w_{i} $. Specifically, for any
    $i,j$, if $ w_{i}=w_{j}=1 $, then the radius set $ R $ is the same as the SB algorithm for
     unweighted network.

\end{itemize}

    In this sense, the SBw algorithm can be applied
to calculate the mass exponents $ \tau(q) $ and the generalized
fractal dimensions $D(q)$ not only for unweighted network but also
for weighted networks. Now we propose a modified SB algorithm (SBw) for
MFA of weighted network as:
\begin{description}
    \item[Step 1] Initially, ensure that all nodes in the network
    are not covered and not selected  as a center of a sandbox.

    \item[Step 2]Set every element in the radius set $R $ as
    the radius $r$ of the sandbox which will be
    used to cover the nodes, where $R$ is obtained as above.
    (in the SB algorithm the radius $r$ in the range $r \in [1,
    d]$, where $d$ is the diameter of the network)

    \item[Step 3] Rearrange the nodes of the entire network into a
      random order. Make sure the nodes of the network are randomly chosen
      as the center of a sandbox.

    \item[Step 4] According to the size $N$ of networks, choose the
        first 1000 nodes in a random order as the center of 1000
       sandboxes, then for each sandbox, search all the neighbor nodes which have a distance
       to the center node within $r$.

    \item[Step 5] Count the number of nodes in each sandbox of radius $r$,
    denote the number of nodes in each sandbox of radius $r$ as $M(r)$.

    \item[Step 6]Calculate the statistical average
         $\langle [M(r)]^{q-1} \rangle$ of $[M(r)]^{q-1}$ over all 1000 sandboxes of radius $r$.

   \item[Step 7]For different values in the radius set $R $ , repeat steps (2) to (6) to
        obtain the statistical average $\langle [M(r)]^{q-1} \rangle$
         and then use $\langle [M(r)]^{q-1} \rangle$ for linear regression.
\end{description}

\section*{Acknowledgement}
~~~ This work is supported by the National Natural Science Foundation of China (Grant No. 11371016), the Chinese Program for Changjiang Scholars and Innovative Research Team in University (PCSIRT) (Grant No. IRT1179), and the Research Foundation of Education Commission of
Hunan Province of China (Grant No. 15C0389).

\vspace{2cm}
\section*{Figures}

\end{document}